\documentclass[pra,aps,twocolumn,showpacs]{revtex4}

\usepackage{amsmath}
\usepackage{bm}
\usepackage{graphicx}

\begin{document}

\title{
Dynamical Casimir effect for magnons in a spinor Bose-Einstein condensate
}

\author{Hiroki Saito$^1$}
\author{Hiroyuki Hyuga$^2$}
\affiliation{$^1$Department of Applied Physics and Chemistry, The
University of Electro-Communications, Tokyo 182-8585, Japan \\
$^2$Department of Physics, Keio University, Yokohama 223-8522, Japan
}

\date{\today}

\begin{abstract}
Magnon excitation in a spinor Bose-Einstein condensate by a driven
magnetic field is shown to have a close analogy with the dynamical Casimir
effect.
A time-dependent external magnetic field amplifies quantum fluctuations in
the magnetic ground state of the condensate, leading to magnetization of
the system.
The magnetization occurs in a direction perpendicular to the magnetic
field breaking the rotation symmetry.
This phenomenon is numerically demonstrated and the excited quantum field
is shown to be squeezed.
\end{abstract}


\maketitle

\section{Introduction}

Vacuum fluctuations play an important role in a variety of situations in
quantum physics.
For instance, the static Casimir effect~\cite{Casimir} originates from
vacuum fluctuations in the electromagnetic field and in the electronic
states in matter.
If the definition of the vacuum state depends on time due to a
time-dependent external condition and the system cannot follow the
instantaneous vacuum state adiabatically, the vacuum fluctuation
materializes as real particles~\cite{Takahashi,Parker,Moore,Fulling}.
This phenomenon is called the nonstationary or dynamical Casimir effect
(DCE).

The DCE has been extensively studied~\cite{Review}, especially for photons
and the massless scalar field.
When the mirror of an optical cavity is rapidly
moved~\cite{Castagnino} or the dielectric constant of the matter in a
cavity is rapidly altered~\cite{Yab,Dodonov93PRA,Johnston,Saito}, photons
are created in the cavity even if the initial state of the electromagnetic
field is in the vacuum state.
The generated photons are in a squeezed state~\cite{Dodonov90}, which
modifies the Casimir force exerted on the mirrors~\cite{Jaekel92}.
The finite temperature correction to the DCE~\cite{Plunien} and
decoherence via the DCE~\cite{Dalvit} have also been studied.
Since there is no intrinsic Hamiltonian in the original formulation of the
moving mirror problem~\cite{Moore}, the effective Hamiltonian approach has
been developed~\cite{Razavy83,Barton,Law94,Haro}.
We have derived an effective Hamiltonian for the moving mirror problem by
quantizing both the electromagnetic field and the polarization field in
the mirrors~\cite{Saito02}.

However, photon creation by the DCE has not yet been experimentally
observed.
This is because the mirror of a cavity must be vibrated at a frequency of
the order of GHz (at least for a microwave cavity) in order to resonantly
amplify the photon field~\cite{Dodonov96}.
Recently, the INFN group~\cite{Braggio} proposed a method to detect the
DCE using a semiconductor layer illuminated by laser pulses, which enables
the rapid displacement of the position of the cavity mirror.
The proposed experiment using this method, however, has not been completed
to date.

In the present paper, we propose a novel system to realize the DCE: a Bose
Einstein condensate (BEC) of an ultracold atomic gas with spin degrees of
freedom.
In this system, magnons are created through the DCE.
The time-dependent external condition that causes the DCE corresponds to
the time-dependent magnetic field applied to the BEC.
Unlike in the case of photons, the typical energy of magnons in a BEC for
a magnetic field of $\sim 1$ G is $\sim h \times 100$ Hz ($h$: Planck
constant).
Magnetic field modulation at this frequency is experimentally feasible.
Continuous amplification of magnons by an oscillating magnetic field
leads to magnetization of the system, which can be observed by {\it in
situ} measurements~\cite{Higbie}.

In a broad sense, quasiparticle excitations in a nonstationary BEC may be
regarded as the DCE.
For example, a time dependent trapping potential~\cite{Castin}, a rapid
increase in the interatomic interaction~\cite{Law02}, and collapse of a
BEC by an attractive interaction~\cite{Calzetta} generate Bogoliubov
quasiparticles.
In these cases, the BEC itself is also excited and its shape depends on
time, whose dynamics is described by the mean-field Gross-Pitaevskii (GP)
equation.
However, it is difficult to distinguish the quasiparticle excitation from
the mean-field excitation, and therefore these systems are unsuitable for
demonstrating the DCE.
In contrast, in our model, the time-dependent magnetic field excites only 
the vacuum fluctuation in the initial quasiparticle vacuum state, giving
an ideal testing ground for the DCE.

This paper is organized as follows.
Section~\ref{s:formulation} formulates the problem.
Section~\ref{s:relation} discusses the relation of the magnon excitation
in the present system with the DCE.
Section~\ref{s:numerical} numerically demonstrates the proposed phenomena 
using the GP equation with quantum fluctuations.
Section~\ref{s:conc} provides discussion and conclusions.

\section{Formulation of the problem}
\label{s:formulation}

\subsection{Hamiltonian for the system}

We consider spin-1 bosonic atoms with mass $M$ confined in an optical
trapping potential $U(\bm{r})$.
The single-particle part of the Hamiltonian without a magnetic field is
given by
\begin{equation}
\hat H_0 = \sum_{m = -1}^1 \int d\bm{r} \hat\psi_m^\dagger(\bm{r}) \left[
-\frac{\hbar^2}{2M} \nabla^2 + U(\bm{r}) \right] \hat\psi_m(\bm{r}),
\end{equation}
where $\hat\psi_m(\bm{r})$ is the field operator that annihilates an atom
with spin magnetic quantum number $m = -1, 0, 1$ at the position
$\bm{r}$.

The interatomic interaction for ultracold spin-1 atoms is described by the
s-wave scattering lengths $a_0$ and $a_2$, where the subscripts 0 and 2
indicate the total spin of two colliding atoms.
The interaction Hamiltonian can be written in spin-independent and
spin-dependent parts as~\cite{Ho,Ohmi}
\begin{equation} \label{Hint}
\hat H_{\rm int} = \int d\bm{r} \left[ c_0 :\hat \rho^2(\bm{r}):
+ c_1 :\hat{\bm{F}}(\bm{r}) \cdot \hat{\bm{F}}(\bm{r}): \right],
\end{equation}
where the symbol $::$ denotes the normal ordering and
\begin{eqnarray}
\label{rho}
\hat \rho(\bm{r}) & = & \sum_{m = -1}^1 \hat\psi_m^\dagger(\bm{r})
\hat\psi_m(\bm{r}), \\
\label{F}
\hat{\bm{F}}(\bm{r}) & = & \sum_{m, m'} \hat\psi_m^\dagger(\bm{r})
(\bm{f})_{m, m'} \hat\psi_{m'}(\bm{r}),
\end{eqnarray}
with $\bm{f} = (f_x, f_y, f_z)$ being the vector of the spin-1 $3 \times
3$ matrices.
The interaction coefficients $c_0$ and $c_1$ in Eq.~(\ref{Hint}) are given
by 
\begin{eqnarray}
c_0 & = & \frac{4 \pi \hbar^2}{M} \frac{a_0 + 2 a_2}{3}, \\
c_1 & = & \frac{4 \pi \hbar^2}{M} \frac{a_2 - a_0}{3}.
\end{eqnarray}

We restrict ourselves to the case of the hyperfine spin $F = 1$ of an
alkali atom with nuclear spin $I = 3/2$ and electron spin $S = 1/2$ (e.g.,
$^{23}{\rm Na}$ and $^{87}{\rm Rb}$).
Because of the hyperfine coupling between the nuclear and electron spins
and their different magnetic moments, the Zeeman energy is a nonlinear
function of $B$~\cite{Pethick}.
Taking the first and second order terms, the Hamiltonian becomes
\begin{equation} \label{HB}
\hat H_B = \int d\bm{r} \left[ p_1 B(t) \hat F_z + p_2 B^2(t)
\left( \hat\psi_1^\dagger \hat\psi_1 + \hat\psi_{-1}^\dagger \hat\psi_{-1}
\right) \right],
\end{equation}
where $p_1$ and $p_2$ are the linear and quadratic Zeeman coefficients,
respectively, and we assume that the uniform magnetic field $B(t)$ is
applied in the $z$ direction.
We define the quadratic Zeeman energy as
\begin{equation}
q(t) \equiv p_2 B^2(t),
\end{equation}
which is positive for the $F = 1$ hyperfine state.

Thus, the total Hamiltonian has the form
\begin{equation} \label{H}
\hat H = \hat H_0 + \hat H_B + \hat H_{\rm int}.
\end{equation}
Since the linear Zeeman term in Eq.~(\ref{HB}) commutes with the other
part of the Hamiltonian and only rotates the spin uniformly around the $z$
axis, we neglect the linear Zeeman term henceforth.

\subsection{Bogoliubov approximation}

The initial state considered in the present paper is the ground state for
an initial value of the quadratic Zeeman energy, $q = q(0)$, under the
restriction
\begin{equation} \label{Fzero}
\int d\bm{r} \langle \hat F_z(\bm{r}) \rangle = 0.
\end{equation}
In Sec.~\ref{s:conc} we discuss how to generate this state.
The spin-independent interaction coefficient $c_0$ must be positive for
the existence of the ground state.
Either for $c_1 > 0$ or for $c_1 < 0$ and $q \gtrsim 2 |c_1| \tilde\rho$,
where $\tilde\rho$ is a typical atomic density, almost all atoms are in
the $m = 0$ state for this initial state~\cite{Stenger}.
We therefore employ the Bogoliubov approximation by setting
\begin{eqnarray}
\label{Bogo0}
\hat \psi_0(\bm{r}) & = & e^{-i \mu t / \hbar} \left[\Psi_0(\bm{r}) +
\hat\phi_0(\bm{r}) \right], \\
\label{Bogo1}
\hat \psi_{\pm 1}(\bm{r}) & = & e^{-i \mu t / \hbar}
\hat\phi_{\pm 1}(\bm{r}),
\end{eqnarray}
where $\Psi_0(\bm{r})$ is a real function that minimizes the energy
functional,
\begin{equation} \label{E0}
E_0 = \int d\bm{r} \left[ \Psi_0 \left( -\frac{\hbar^2}{2M} \nabla^2 + U
\right) \Psi_0 + \frac{c_0}{2} \Psi_0^4 \right],
\end{equation}
$\mu$ is the chemical potential,
\begin{equation}
\mu = \int d\bm{r} \left[ \Psi_0 \left( -\frac{\hbar^2}{2M} \nabla^2 + U
\right) \Psi_0 + c_0 \Psi_0^4 \right],
\end{equation}
and $\hat\phi_{m}(\bm{r})$ are the fluctuation operators.
The wave function $\Psi_0(\bm{r})$ is normalized as $\int d\bm{r}
\Psi_0^2(\bm{r}) = N$ with $N$ being the number of atoms.
The Heisenberg equations of motion for $\hat\phi_m$ are then written
as~\cite{Ho}
\begin{eqnarray}
i \hbar \frac{\partial \hat\phi_0}{\partial t} & = & \left(
-\frac{\hbar^2}{2 M} \nabla^2 + U - \mu \right) \hat\phi_0
+ c_0 \Psi_0^2 \left( 2\hat\phi_0 + \hat\phi_0^\dagger \right),
\nonumber \\
\label{phi0eq}
\\
\label{phi1eq}
i \hbar \frac{\partial \hat\phi_{\pm 1}}{\partial t} & = & \left(
-\frac{\hbar^2}{2 M} \nabla^2 + U + q + c_0 \Psi_0^2 - \mu \right)
\hat\phi_{\pm 1} \nonumber \\
& & + c_1 \Psi_0^2 \left( \hat\phi_{\pm 1} + \hat\phi_{\mp
1}^\dagger \right),
\end{eqnarray}
where we neglect the second and third orders of $\hat\phi_m$.
Equation~(\ref{phi0eq}) is identical with that of a scalar BEC, which
gives the Bogoliubov spectrum $E^{(0)}_k = \sqrt{\varepsilon_{k}
(\varepsilon_{k} + 2 c_0 \Psi_0^2)}$ for a uniform system, where
$\varepsilon_{k} = \hbar^2 k^2 / (2 M)$.

If $q$ is constant in time, we can solve Eq.~(\ref{phi1eq}) by setting
\begin{equation} \label{phi1exp}
\hat\phi_{\pm 1} = \sum_\lambda \left[ u_\lambda(\bm{r}) \hat b_{\pm,
\lambda} e^{-i E^{(1)}_\lambda t / \hbar} + v_\lambda^*(\bm{r}) \hat
b_{\mp, \lambda}^\dagger e^{i E^{(1)}_\lambda t / \hbar} \right], 
\end{equation}
where $\hat b_{\pm, \lambda}$ are bosonic operators for quasiparticles and
$u_\lambda$, $v_\lambda$, and $E^{(1)}_\lambda$ are determined by the
Bogoliubov-de Genne equations,
\begin{eqnarray}
& & \left( -\frac{\hbar^2}{2 M} \nabla^2 + U + q + c_0 \Psi_0^2 + c_1
\Psi_0^2 - \mu \right) u_\lambda + c_1 \Psi_0^2 v_\lambda
\nonumber \\
& = & E^{(1)}_\lambda u_\lambda,
\label{BdG1}
\\
& & \left( -\frac{\hbar^2}{2 M} \nabla^2 + U + q + c_0 \Psi_0^2 + c_1
\Psi_0^2 - \mu \right) v_\lambda + c_1 \Psi_0^2 u_\lambda
\nonumber \\
& = & -E^{(1)}_\lambda v_\lambda.
\label{BdG2}
\end{eqnarray}
The quasiparticles created by $\hat b_{\pm, \lambda}^\dagger$ are
excitations of $m = \pm 1$ states, which we call magnons.

For a uniform system, Eqs.~(\ref{BdG1}) and (\ref{BdG2}) are solved to
give
\begin{eqnarray}
u_{k} & = & \frac{e^{i \bm{k} \cdot \bm{r}}}{\sqrt{2V}}
\sqrt{\frac{\varepsilon_{k} + q + c_1 \Psi_0^2}{E^{(1)}_k} + 1}, \\
v_{k} & = & -\frac{e^{i \bm{k} \cdot \bm{r}}}{\sqrt{2V}}
\sqrt{\frac{\varepsilon_{k} + q + c_1 \Psi_0^2}{E^{(1)}_k} - 1},
\end{eqnarray}
where $V$ is the volume of the system and
\begin{equation} \label{E1k}
E^{(1)}_k = \sqrt{(\varepsilon_{k} + q) (\varepsilon_{k} + q + 2 c_1
\Psi_0^2)}.
\end{equation}
When all Bogoliubov energies are real, the system is dynamically stable.
If $c_1 < 0$ and $q < 2 |c_1| \Psi_0^2$, Eq.~(\ref{E1k}) is imaginary for
long wavelengths and the system is dynamically unstable against
spontaneous magnetization~\cite{Saito05,SaitoL}.
In the present paper, we consider such parameters that the system is
dynamically stable.

\subsection{Effective Hamiltonian for oscillating $q$}

Here we assume that the quadratic Zeeman energy $q(t)$ oscillates as
\begin{equation} \label{q}
q(t) = q_0 (1 + \delta_q \sin \Omega t)
\end{equation}
through the oscillation of the strength of the applied magnetic field.
For $\delta_q = 0$, the Bogoliubov Hamiltonian for magnons is given by
\begin{equation}
\hat H_{\rm mag}^{q = q_0} = \sum_\lambda E^{(1)}_\lambda \left( \hat
b_{+, \lambda}^\dagger \hat b_{+, \lambda} + \hat b_{-, \lambda}^\dagger
\hat b_{-, \lambda} \right),
\end{equation}
where $E^{(1)}_\lambda$, $\hat b_{+, \lambda}$, and $\hat b_{-, \lambda}$
are energies and annihilation operators of magnons defined at $q = q_0$.
The magnon Hamiltonian for $\delta_q \neq 0$ is then written as
\begin{equation} \label{Hmag}
\hat H_{\rm mag}^q = \hat H_{\rm mag}^{q = q_0} + q_0 \delta_q \sin
\Omega t \int d\bm{r} \left( \hat\phi_1^\dagger \hat\phi_1 +
\hat\phi_{-1}^\dagger \hat\phi_{-1} \right).
\end{equation}
When the last term in Eq.~(\ref{Hmag}) is treated as the perturbation, its
interaction representation becomes
\begin{eqnarray} \label{Heff1}
& & q_0 \delta_q \sin \Omega t \sum_{\lambda \lambda'} \int d\bm{r}
\Bigl[ e^{i (E^{(1)}_\lambda - E^{(1)}_{\lambda'}) t / \hbar}
\left( u_\lambda^* u_{\lambda'} + v_\lambda^* v_{\lambda'} \right)
\nonumber \\
& & \times \left( \hat b_{+, \lambda}^\dagger \hat b_{+, \lambda'}
+ \hat b_{-, \lambda}^\dagger \hat b_{-, \lambda'} \right)
\nonumber \\
& & + e^{-i (E^{(1)}_\lambda + E^{(1)}_{\lambda'}) t / \hbar}
\left( u_\lambda v_{\lambda'} + v_\lambda u_{\lambda'} \right)
\hat b_{+, \lambda} \hat b_{-, \lambda'}
\nonumber \\
& & + {\rm H. c.} + {\rm const.}
\Bigr],
\end{eqnarray}
where ${\rm H. c.}$ denotes the Hermitian conjugate of the preceding
term.
If the frequency $\Omega$ is resonant with $(E^{(1)}_\lambda +
E^{(1)}_{\lambda'}) / \hbar$ for some $\lambda$ and $\lambda'$, the terms
proportional to $\hat b_{+, \lambda} \hat b_{-, \lambda'}$ and $\hat b_{+,
\lambda}^\dagger \hat b_{-, \lambda'}^\dagger$ become dominant. 

When the system is uniform and $\Omega$ is resonant with $2 E^{(1)}_k /
\hbar$, the Hamiltonian in Eq.~(\ref{Heff1}) reduces to
\begin{equation} \label{Heff}
\hat H_{\rm eff} = -i q_0 \delta_q V {\sum_{\bm{k}}}' |u_k v_k|
\left( \hat b_{+, \bm{k}} \hat b_{-, -\bm{k}} - 
\hat b_{+, \bm{k}}^\dagger \hat b_{-, -\bm{k}}^\dagger \right),
\end{equation}
where ${\sum_{\bm{k}}}'$ denotes that the summation is taken for $\hbar^2
k^2 / (2 M) \simeq 2 E^{(1)}_k$.
The form of Eq.~(\ref{Heff}) suggests that the magnon state develops into
the squeezed state by the oscillation of $q$, which will be confirmed in
Sec.~\ref{s:num2} numerically.

\section{Relation with the dynamical Casimir effect}
\label{s:relation}

We now discuss the relation of the present system to the usual DCE.

First let us review the DCE for an electromagnetic field.
The initial state is the vacuum state for some static configurations of
mirrors and dielectrics.
The initial expectation values of the electric and magnetic fields
therefore vanish: $\langle \hat{\bm{E}} \rangle = \langle \hat{\bm{H}}
\rangle = 0$.
These operators obey the Heisenberg equations of motion,
\begin{eqnarray}
\label{Eeq}
\frac{\partial}{\partial t} \varepsilon(\bm{r}, t) \hat{\bm{E}}(\bm{r}, t)
& = & \bm{\nabla} \times \hat{\bm{H}}(\bm{r}, t), \\
\label{Heq}
\frac{\partial}{\partial t} \hat{\bm{H}}(\bm{r}, t)
& = & -\bm{\nabla} \times \varepsilon(\bm{r}, t) \hat{\bm{E}}(\bm{r}, t),
\end{eqnarray}
where $\varepsilon$ is a dielectric constant, and the electric field
operator $\hat{\bm{E}}$ must vanish at the mirrors.
From Eqs.~(\ref{Eeq}) and (\ref{Heq}), $\langle \hat{\bm{E}} \rangle$ and
$\langle \hat{\bm{H}} \rangle$ always remain zero as in classical
electrodynamics, whereas the vacuum fluctuation of the electromagnetic
field can be amplified, leading to the creation of photons.

For the present spinor BEC system, the initial state is assumed to be the
ground state satisfying Eq.~(\ref{Fzero}).
Since the Hamiltonian (\ref{H}) has spin-rotation symmetry around the $z$
axis, the expectation value of the transverse magnetization for the
initial state vanishes: $\langle \hat F_+ \rangle = 0$, where $\hat F_+ =
\hat F_x + i \hat F_y$.
In the Bogoliubov approximation, using Eqs.~(\ref{Bogo0}) and
(\ref{Bogo1}), the magnetization operator $\hat F_+ = \sqrt{2} (\hat
\psi_1^\dagger \hat \psi_0 + \hat \psi_0^\dagger \hat \psi_{-1})$ reduces
to
\begin{equation} \label{Fpls}
\hat F_+ \simeq \sqrt{2} \Psi_0 \left( \hat\phi_1^\dagger + \hat\phi_{-1}
\right).
\end{equation}
From Eq.~(\ref{phi1eq}), the Heisenberg equations of motion for $\hat F_+$
are obtained as
\begin{eqnarray}
\hbar \frac{\partial \hat F_+}{\partial t} & = & \left[
-\frac{\hbar^2}{2M} \nabla^2 + V + q(t) + c_0 \Psi_0^2 - \mu \right]
\hat \Pi_+,
\label{Feq}
\\
\hbar \frac{\partial \hat \Pi_+}{\partial t} & = & -\biggl[
-\frac{\hbar^2}{2M} \nabla^2 + V + q(t)
\nonumber \\
& & + (c_0 + 2 c_1) \Psi_0^2 - \mu \biggr] \hat F_+,
\label{Pieq}
\end{eqnarray}
where we define
\begin{equation}
\hat \Pi_+ = -i \sqrt{2} \Psi_0 \left( \hat\phi_1^\dagger -
\hat\phi_{-1} \right).
\end{equation}
The expectation value of this operator also vanishes, $\langle \hat\Pi_+
\rangle = 0$, for the initial state.
We find from Eqs.~(\ref{Feq}) and (\ref{Pieq}) that $\langle \hat F_+
\rangle$ and $\langle \hat\Pi_+ \rangle$ always remain zero.
On the other hand, their quantum fluctuations can be amplified due to the
temporal variation of $q(t)$, leading to the creation of Bogoliubov
quasiparticles, i.e., magnons.
Thus, the present situation is similar to that of the DCE in the
electromagnetic field in that the amplification of quantum fluctuations
and (quasi)particle creations occur while the expectation values of the
fields remain constant ($\langle \hat F_+ \rangle = \langle \hat\Pi_+
\rangle = 0$).

What kind of excitation phenomenon can we regard as the DCE?
For example, can the excitation of ripples on water in a vibrating bucket
be attributed to the DCE?
The answer is partially yes, because amplification of the quantum
fluctuation on the water surface does occur, which creates ripplons.
However, this quantum excitation is overwhelmed by the classical
excitation of the water surface and identification of the quasiparticle
excitations should therefore be extremely difficult.
Thus, the suitable condition for studying the DCE is that only the quantum
fluctuation is excited and the classical field, i.e., the expectation
value of the relevant quantum field, remains constant during the temporal
variation of external parameters.
For the electromagnetic field, $\langle \hat{\bm{E}}(\bm{r}, t) \rangle =
\langle \hat{\bm{H}}(\bm{r}, t) \rangle = 0$, even when the position of
the mirror and $\varepsilon(\bm{r}, t)$ of matter are changed.
The present spinor BEC system is also suitable for studying the DCE, since
$\langle \hat F_+(\bm{r}, t) \rangle = \langle \hat\Pi_+(\bm{r}, t)
\rangle = 0$ holds during the change of $q(t)$.

\section{Numerical analysis}
\label{s:numerical}

\subsection{Mean-field theory with quantum fluctuations}

In this section, we numerically demonstrate the DCE of magnons in a spinor
BEC.
Since performing a full quantum many-body simulation is difficult, we
employ a mean-field approximation taking into account the initial quantum
fluctuations.
Recently, this method was used to predict spin vortex formation through
the Kibble-Zurek mechanism in magnetization of a spinor
BEC~\cite{SaitoKZ}.

The mean-field GP equations for a spin-1 BEC are given by
\begin{eqnarray}
i \hbar \frac{\partial \psi_{\pm 1}}{\partial t} & = & \left[
-\frac{\hbar^2}{2 M} \nabla^2 + V + q(t) + c_0 \rho \right] \psi_{\pm 1}
\nonumber \\
& & + c_1 \left( \frac{1}{\sqrt{2}} F_\mp \psi_0 \pm F_z  \psi_{\pm 1}
\right),
\label{GP1} \\
i \hbar \frac{\partial \psi_0}{\partial t} & = & \left( -\frac{\hbar^2}{2
M} \nabla^2 + V + c_0 \rho \right) \psi_0
\nonumber \\
& & + \frac{c_1}{\sqrt{2}} \left(  F_+ \psi_1 + F_- \psi_{-1} \right),
\label{GP2}
\end{eqnarray}
where $\psi_m(\bm{r}, t)$ are the macroscopic wave functions and $\rho$,
$F_z$, and $F_\pm = F_x \pm i F_y$ are defined by the forms in
Eqs.~(\ref{rho}) and (\ref{F}) in which $\hat\psi_m$ are replaced by
$\psi_m$.
The initial wave function for $\psi_0$ is the ground state solution
$\Psi_0$ of Eq.~(\ref{E0}), which is a stationary solution of
Eq.~(\ref{GP2}) for $\psi_{\pm 1} = 0$.
Although $\langle \hat F_\pm \rangle = 0$ for the exact many-body initial
state, we do not set $\psi_{\pm 1} = 0$, since the right-hand side of
Eq.~(\ref{GP1}) vanishes and no time evolution is obtained.

We include a small initial noise in $\psi_{\pm 1}$ to reproduce vacuum
fluctuation in magnetization within the Bogoliubov approximation.
To this end, we consider the magnetic correlation function for the
Bogoliubov vacuum,
\begin{equation} \label{FF}
\langle \hat F_+(\bm{r}) \hat F_-(\bm{r}') \rangle \simeq 2 \Psi_0(\bm{r})
\Psi_0(\bm{r}') \sum_\lambda f_\lambda(\bm{r}) f_\lambda^*(\bm{r}'),
\end{equation}
where $f_\lambda(\bm{r}) \equiv u_\lambda(\bm{r}) + v_\lambda(\bm{r})$ and
we have used Eqs.~(\ref{phi1exp}) and (\ref{Fpls}).
To find the appropriate mean-field initial states of $\psi_{\pm 1}$, we
assume the form
\begin{equation} \label{psi1ini}
\psi_{\pm 1}(\bm{r}) = \sum_\lambda \left[ u_\lambda(\bm{r}) b_{\pm,
\lambda} + v_\lambda^*(\bm{r}) b_{\mp, \lambda}^* \right],
\end{equation}
where $b_{\pm, \lambda}$ are random numbers whose probability distribution
is determined below.
Substituting Eq.~(\ref{psi1ini}) into $F_+(\bm{r}) F_-(\bm{r}')$ and
taking the average with respect to the probability distribution, we obtain
\begin{eqnarray} \label{FF2}
\langle F_+(\bm{r}) F_-(\bm{r}') \rangle_{\rm avg} & = & 2 \Psi_0(\bm{r})
\Psi_0(\bm{r}') 
\nonumber \\
& & \times \sum_{\lambda, \lambda'} \bigl[ f_\lambda^*(\bm{r})
f_{\lambda'}(\bm{r}') \langle b_{+, \lambda}^* b_{+, \lambda'}
\rangle_{\rm avg}
\nonumber \\
& & + f_\lambda(\bm{r}) f_{\lambda'}^*(\bm{r}')
\langle b_{-, \lambda} b_{-, \lambda'}^* \rangle_{\rm avg}
\nonumber \\
& & + f_\lambda(\bm{r}) f_{\lambda'}(\bm{r}')
\langle b_{-, \lambda} b_{+, \lambda'} \rangle_{\rm avg}
\nonumber \\
& & + f_\lambda^*(\bm{r}) f_{\lambda'}^*(\bm{r}')
\langle b_{+, \lambda}^* b_{-, \lambda'}^* \rangle_{\rm avg} \bigr].
\end{eqnarray}
Using the fact that both $(u, v)$ and $(u^*, v^*)$ are solutions of
Eqs.~(\ref{BdG1}) and (\ref{BdG2}) and assuming that the random variables
$b_{+, \lambda}$ and $b_{-, \lambda}$ obey the same distribution, we find
that Eq.~(\ref{FF2}) coincides with Eq.~(\ref{FF}) by a probability
distribution satisfying
\begin{equation} \label{bdist}
\langle b_{\pm, \lambda}^* b_{\pm, \lambda'} \rangle_{\rm avg} = 
\frac{1}{2} \delta_{\lambda, \lambda'}, \;\;\;\;\;\;
\langle b_{\pm, \lambda} b_{\mp, \lambda'} \rangle_{\rm avg} = 0.
\end{equation}
We also assume $\langle b_{\pm, \lambda} b_{\pm, \lambda'}
\rangle_{\rm avg} = \langle b_{\pm, \lambda} b_{\mp, \lambda'}^*
\rangle_{\rm avg} = 0$.
Thus, the initial mean-field states (\ref{psi1ini}) with probability
distributions obeying Eq.~(\ref{bdist}) reproduce the same magnetic
correlation as for the Bogoliubov approximation.
Similarly, we can show that the time evolution of $\langle F_+(\bm{r})
F_-(\bm{r}') \rangle_{\rm avg}$ also agrees with that of $\langle \hat
F_+(\bm{r}) \hat F_-(\bm{r}') \rangle$ at the level of the Bogoliubov
approximation.

\subsection{Numerical results}
\label{s:num2}

We numerically solve Eqs.~(\ref{GP1}) and (\ref{GP2}) for the
time-dependent magnetic field that gives a sinusoidal oscillation of the
quadratic Zeeman energy as in Eq.~(\ref{q}).
For simplicity, we consider a uniform two-dimensional (2D) system with a
periodic boundary condition.
We assume that $10^7$ $^{23}{\rm Na}$ atoms are in a $100$ $\mu{\rm m}
\times 100$ $\mu{\rm m}$ space with atomic density $n = 2.8 \times
10^{14}$ ${\rm cm}^{-3}$.
In a recent experiment~\cite{Black}, $a_2 - a_0 \propto c_1$ for an $F =
1$ $^{23}{\rm Na}$ atom was precisely measured to be about 2.47 Bohr
radii.
The system is therefore dynamically stable according to Eq.~(\ref{E1k}).
The initial state is $\psi_0 = \sqrt{n}$ and $\psi_{\pm 1}$ are given by
Eqs.~(\ref{psi1ini}) and (\ref{bdist}) in which the complex random
variables $b_{\pm, \bm{k}}$ are assumed to follow the Gaussian
distribution $P(b) = \sqrt{2 / \pi} e^{-2 |b|^2}$.
The initial random variables $b_{\pm, \bm{k}}$ are cut off for $2\pi / k
\lesssim 1.5$ $\mu{\rm m}$.
The strength of the magnetic field at $t = 0$ is 500 mG, which corresponds
to $q_0 \simeq h \times 69$ Hz $\simeq 1.02 c_1 n$.

\begin{figure}[t]
\includegraphics[width=8cm]{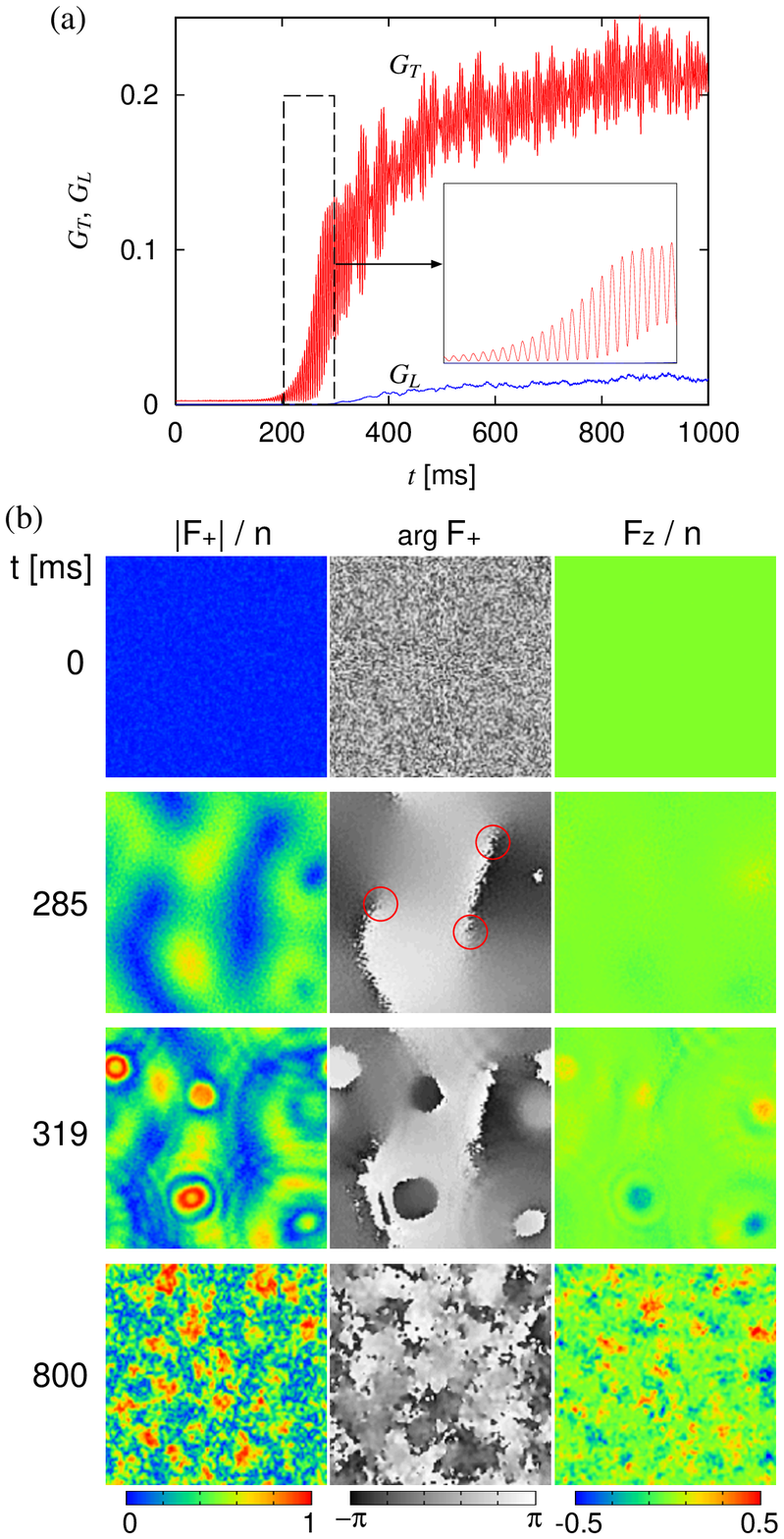}
\caption{
(Color) (a) Time evolution of the average squared transverse magnetization
$G_{\rm T}$ (red curve) and longitudinal magnetization $G_{\rm L}$ (blue
curve) for $n = 2.8 \times 10^{14}$ ${\rm cm}^{-3}$, $N = 10^7$, 
$\Omega = 2 \sqrt{q_0 (q_0 + 2 c_1 n)} \simeq 2 \pi \times 237$ Hz, and
$\delta_q = 0.2$.
The inset magnifies the dashed square.
(b) Profiles of $|F_+| / n$, ${\rm arg} F_+$, and $F_z / n$ at $t =
0$, 285, 319, and 800 ms.
The field of view is $100$ $\mu{\rm m} \times 100$ $\mu{\rm m}$.
The red circles indicate the topological defects.
}
\label{f:ev}
\end{figure}
Figure~\ref{f:ev} (a) shows time evolution of the average squared
transverse magnetization~\cite{Sadler},
\begin{equation}
G_{\rm T} = \frac{\int d\bm{r} |F_+(\bm{r})|^2}{\int d\bm{r}
\rho^2(\bm{r})},
\end{equation}
and longitudinal magnetization,
\begin{equation}
G_{\rm L} = \frac{\int d\bm{r} |F_z(\bm{r})|^2}{\int d\bm{r}
\rho^2(\bm{r})},
\end{equation}
for $\Omega = 2 \sqrt{q_0 (q_0 + 2 c_1 n)} \simeq 2 \pi \times 237$ Hz and
$\delta_q = 0.2$.
This frequency is resonant with $2 E^{(1)}_{k = 0} / \hbar$, and the $k
\simeq 0$ modes are expected to be excited according to the effective
Hamiltonian (\ref{Heff}).
From Fig.~\ref{f:ev} (a), we find that the time constant for the
exponential growth of $G_{\rm T}$ is $\simeq 22$ ms.

Solving the Heisenberg equation for the effective Hamiltonian with $k =
0$,
\begin{equation}
\hat H_{\rm eff}^0 = -i \xi_0 \left( \hat b_{+, 0} \hat b_{-, 0} - \hat
b_{+, 0}^\dagger \hat b_{-, 0}^\dagger \right),
\end{equation}
we obtain
\begin{equation}
\hat b_{\pm, 0}(t) = \hat b_{\pm, 0} e^{-i E^{(1)}_0 t / \hbar} \cosh
\xi_0 t + \hat b_{\mp, 0}^\dagger e^{i E^{(1)}_0 t / \hbar} \sinh \xi_0 t,
\end{equation}
where $\xi_0 \equiv q_0 \delta_q |u_0 v_0| V$.
Substitution of this solution into Eq.~(\ref{Fpls}) with
\begin{equation} \label{phi1}
\hat\phi_{\pm 1}(\bm{r}, t) = \sum_{\bm{k}} \left[ u_k(\bm{r})
\hat b_{\pm, \bm{k}}(t) + v_k^*(\bm{r}) \hat b_{\pm, \bm{k}}^\dagger(t)
\right]
\end{equation}
gives
\begin{eqnarray} \label{Fgrow}
\hat F_+(t) & \simeq & \sqrt{2n} \sum_{\bm{k}} e^{-i \bm{k} \cdot
\bm{r}} \left( |u_k| + |v_k| \right) \left[ \hat b_{+, \bm{k}}^\dagger(t)
+ \hat b_{-, -\bm{k}}(t) \right]
\nonumber \\
& \simeq & \sqrt{2n} \left( |u_0| + |v_0| \right)
\nonumber \\
& & \times \left( \hat b_{+, 0}^\dagger e^{i E^{(1)}_0 t / \hbar} + \hat
b_{-, 0} e^{-i E^{(1)}_0 t / \hbar} \right) e^{\xi_0 t},
\end{eqnarray}
where in the second line we take only the $k = 0$ component because of the
exponential factor $e^{\xi_0 t}$.
The average squared transverse magnetization $G_{\rm T}$ thus grows
exponentially with a time constant $1 / (2 \xi_0) \simeq 20$ ms, which is
in good agreement with the numerical results of 22 ms.

The spatial profile of the magnetization is shown in Fig.~\ref{f:ev} (b).
The transverse magnetization first grows with long wavelength [the second
row of Fig.~\ref{f:ev} (b)], in which we can see a few topological
defects (red circles).
These spin vortices are generated through the Kibble-Zurek
mechanism~\cite{SaitoKZ,Damski}.
The transverse magnetization then exhibits an interesting concentric
pattern [the third row of Fig.~\ref{f:ev} (b)], which may be due to the
nonlinearity.
The magnetization finally breaks into complicated fragments and the
longitudinal magnetization also begins to grow [the fourth row of
Fig.~\ref{f:ev} (b)].

We now discuss symmetry breaking in the DCE.
The original Hamiltonian (\ref{H}) commutes with $\int \hat F_z d\bm{r}$
and the system has spin-rotation symmetry around the $z$ axis.
In reality, however, local magnetization in the $x$-$y$ directions occurs
in a ferromagnetic BEC breaking the spin-rotation symmetry
spontaneously~\cite{Sadler}, and we expect that the symmetry breaking
also occurs in the present system.
This is because the exact quantum state with spin-rotation symmetry is
the macroscopic superposition of magnetized states in the $x$-$y$
directions, which is therefore fragile against external perturbations.
Such symmetry breaking phenomena should also occur in the DCE of the
electromagnetic field.
For example, in the 1D moving mirror problem~\cite{Castagnino}, there is
rotation symmetry around the axis of the 1D cavity and the generated
photon state is a superposition of all polarizations, giving $\langle
\hat{\bm{E}} \rangle = \langle \hat{\bm{B}} \rangle = 0$.
When we measure the local electromagnetic field, however, we obtain a
nonzero value as a result of the symmetry breaking.

\begin{figure}[t]
\includegraphics[width=8cm]{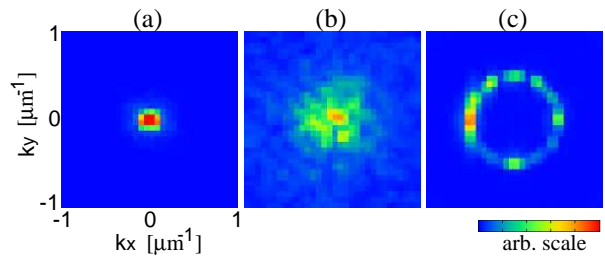}
\caption{
(Color) Magnitude of the Fourier transform of $F_+$ at (a) $t = 285$
ms and (b) $t = 800$ ms.
The parameters are the same as in Fig.~\ref{f:ev}.
(c) Magnitude of the Fourier transform of $F_+$ at $t = 500$ ms for
$\Omega = 2\pi \times 358$ Hz.
The other parameters are the same as in Fig.~\ref{f:ev}.
}
\label{f:fourier}
\end{figure}
Figure~\ref{f:fourier} shows the magnitude of the Fourier transform of
$F_+$,
\begin{equation}
\left| \tilde{F_+}(\bm{k}) \right| = \left| \int d\bm{r}
F_+(\bm{r}) \frac{e^{i \bm{k} \cdot \bm{r}}}{\sqrt{V}} \right|.
\end{equation}
Initially only the $k \simeq 0$ modes grow as shown in
Fig.~\ref{f:fourier} (a).
Then, the momentum distribution becomes broad as shown in
Fig.~\ref{f:fourier} (b).
Specific wave numbers can be selectively excited using larger $\Omega$.
Figure~\ref{f:fourier} (c) shows the result for $\Omega = 2\pi \times 358$
Hz, which is resonant with $2 E^{(1)}_k / \hbar$ for $k = 0.5$ $\mu{\rm
m}^{-1}$.
We can see the ring at this wave number.

From the form of the effective Hamiltonian in Eq.~(\ref{Heff}), generated
magnons are expected to have quantum correlations.
Defining new operators,
\begin{eqnarray}
\label{Bplus}
\hat B_{+, \bm{k}} & = & \frac{1}{\sqrt{2}} \left( \hat b_{+, \bm{k}} +
\hat b_{-, \bm{k}} \right), \\
\hat B_{-, \bm{k}} & = & \frac{1}{\sqrt{2}} \left( \hat b_{+, \bm{k}} -
\hat b_{-, \bm{k}} \right),
\end{eqnarray}
we can rewrite Eq.~(\ref{Heff}) as
\begin{equation}
\hat H_{\rm eff} = -\frac{i}{2} {\sum_{\bm{k}}}' \xi_k \left(
\hat B_{+, \bm{k}}^2 - \hat B_{+, \bm{k}}^{\dagger 2} -
\hat B_{-, \bm{k}}^2 + \hat B_{-, \bm{k}}^{\dagger 2} \right),
\end{equation}
where $\xi_k \equiv q_0 \delta_q |u_k v_k| V$.
We can clearly see that this Hamiltonian generates the squeezed state.

\begin{figure}[t]
\includegraphics[width=8cm]{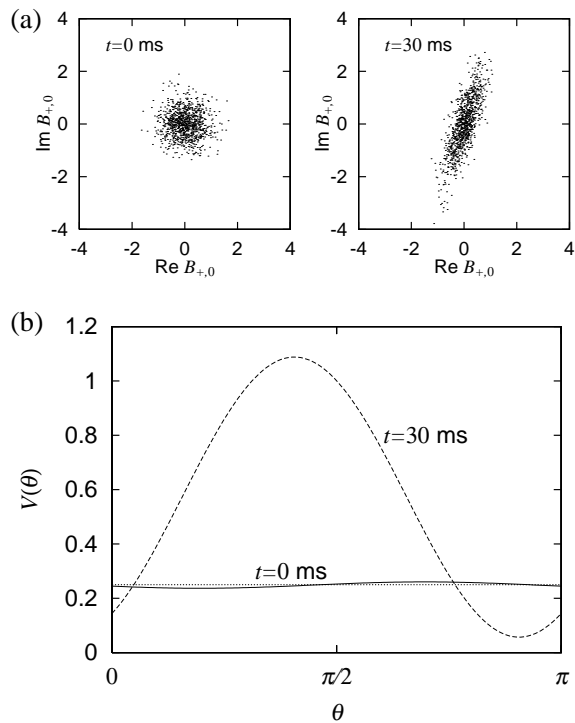}
\caption{
(a) Distributions of $B_{+, 0} = (b_{+, 0} + b_{-, 0}) / \sqrt{2}$ at $t =
0$ and 30 ms obtained by 1000 simulation runs with different initial
states produced by random numbers.
The parameters are the same as in Fig.~\ref{f:ev}.
(b) Variance $V(\theta)$ of ${\rm Re}(B_{+, 0} e^{-i \theta})$ for the
data in (a).
The dotted line indicates $V(\theta) = 1 / 4$.
}
\label{f:squeeze}
\end{figure}
Figure~\ref{f:squeeze} (a) plots the values of $B_{+, 0} = (b_{+, 0} +
b_{-, 0}) / \sqrt{2}$ at $t = 0$ and 30 ms.
The distribution of $B_{+, 0}$ corresponds to the quantum fluctuation in
Eq.~(\ref{Bplus}) with $\bm{k} = 0$, which is expected to be squeezed.
The values of $b_{\pm, 0}$ are obtained from $b_{\pm, \bm{k}} = \sqrt{V}
[|u_k| \tilde\psi_{\pm 1}(\bm{k}) - |v_k| \tilde\psi_{\mp 1}^*(\bm{k})]$
with $\tilde\psi_{\pm 1}(\bm{k})$ being the Fourier transform of
$\psi_{\pm 1}(\bm{r})$.
Each point in Fig.~\ref{f:squeeze} (a) corresponds to a single simulation
run and 1000 simulations are performed.
Figure~\ref{f:squeeze} (b) shows the variance of ${\rm Re}(B_{+, 0} e^{-i
\theta})$,
\begin{equation}
V(\theta) = \langle [{\rm Re}(B_{+, 0} e^{-i \theta})]^2 \rangle_{\rm avg}
- \langle {\rm Re}(B_{+, 0} e^{-i \theta}) \rangle_{\rm avg}^2.
\end{equation}
For the initial state, the variance is isotropic, $V(\theta) = 1 / 4$,
from Eq.~(\ref{bdist}).
The small deviation of the solid curve from $1 / 4$ in
Fig.~\ref{f:squeeze} (b) is the statistical error.
At $t = 30$ ms, the distribution of $B_{+, 0}$ is clearly squeezed, and
the maximum and minimum values of $V(\theta)$ are 1.09 and 0.057.
For the present parameters, $e^{2\xi t} \simeq 4.38$ at $t = 30$ ms, which
is in good agreement with $1.09 / 0.25 \simeq 0.25 / 0.057$.
Since $1.09 \times 0.057 \simeq 1 / 4^2$, the squeezed state is almost the
minimum uncertainty state.

In this section, we assumed the use of $^{23}{\rm Na}$ atoms.
Similar results can be obtained also for $F = 1$ $^{87}{\rm Rb}$ atoms
with a ferromagnetic interaction ($c_1 < 0$), where $q(t)$ must always be
larger than $|c_1| \Psi_0^2$ in order to suppress the spontaneous
magnetization.

\section{Discussion and conclusions}
\label{s:conc}

We now discuss the possibility of experimental observation of the proposed
phenomena.
The magnetization profile $\bm{F}(\bm{r})$ as shown in Fig.~\ref{f:ev} can
be measured by spin-sensitive phase-contrast imaging in a nondestructive
manner~\cite{Higbie,Sadler}, from which $G_{\rm T}$, $G_{\rm L}$, and
$\tilde{\bm{F}}(\bm{k})$ are obtained.
The squeezing of the field as shown in Fig.~\ref{f:squeeze} can also be
observed.
From Eq.~(\ref{Fgrow}), we can measure both $b_{+, \bm{k}}^* + b_{-,
-\bm{k}}$ and $b_{+, \bm{k}}^* - b_{-, -\bm{k}}$, and therefore we obtain
$B_{\pm, \bm{k}}$.
In order to assure that the observed magnetization is definitely due to
amplification of the vacuum fluctuation, we must prepare the appropriate
initial state.
In principle, the initial magnon vacuum state can be prepared as follows.
First, a BEC is prepared in the $m = 0$ state at a sufficiently strong
magnetic field ($q \gg c_1 \Psi_0^2$), in which the $m = 0$ state is
almost the ground state.
The residual atoms in the $m = \pm 1$ states must be eliminated
completely.
Then, the magnetic field is adiabatically decreased to the desired
strength, which gives a magnon vacuum state satisfying Eq.~(\ref{Fzero}).

In conclusion, we have studied the magnon excitation in a spinor BEC by a
driven external magnetic field and have demonstrated the close analogy of
this phenomenon with the DCE.
The present system is suitable for studying the DCE of quasiparticles,
since the time-dependent magnetic field applied to the Bogoliubov ground
state amplifies only the vacuum fluctuation, keeping the ``classical
fields'' constant in the Bogoliubov approximation.

We numerically demonstrated magnon excitation in a spinor BEC using the
mean-field GP equation, in which the vacuum fluctuation is taken into
account by the initial random noise.
We have shown that the oscillating external magnetic field resonantly
amplifies the vacuum fluctuation, leading to magnetization of the
system (Fig.~\ref{f:ev}).
The Fourier transform of the excited field reveals that the specific wave
number $k$ can be selectively amplified (Fig.~\ref{f:fourier}).
The excited quantum field is squeezed (Fig.~\ref{f:squeeze}) as in the DCE
of photons.
The growth of magnetization in Fig.~\ref{f:ev} (a) and the degree of
squeezing in Fig.~\ref{f:squeeze} can be well described by the effective
Hamiltonian in Eq.~(\ref{Heff}).

A spinor BEC is thus a good testing ground for the DCE and our proposal
is feasible with current experimental techniques.
Study of the DCE of quasiparticles may serve as a stepping stone to the
observation of the DCE of photons.

\begin{acknowledgments}
This work was supported by the Ministry of Education, Culture, Sports,
Science and Technology of Japan (Grants-in-Aid for Scientific Research,
No.\ 17071005 and No.\ 20540388) and by the Matsuo Foundation.
\end{acknowledgments}


\begin{thebibliography}{99}

\bibitem{Casimir}
H. B. G. Casimir, Proc. K. Ned. Akad. Wet. {\bf 51}, 793 (1948).

\bibitem{Takahashi}
Y. Takahashi and H. Umezawa, Nuovo Cimento {\bf 6}, 1324 (1957).

\bibitem{Parker}
L. Parker, Phys. Rev. Lett. {\bf 21}, 562 (1968).

\bibitem{Moore}
G. T. Moore, J. Math. Phys. {\bf 11}, 2679 (1970).

\bibitem{Fulling}
S. A. Fulling and P. C. W. Davies, Proc. R. Soc. London A {\bf 348}, 393
(1976).

\bibitem{Review}
For review, see, V. V. Dodonov, Adv. Chem. Phys. {\bf 119}, 309 (2001);
V. V. Dodonov and A. V. Dodonov, J. Phys.: Conf. Ser. {\bf 99}, 012006
(2008).

\bibitem{Castagnino}
See, e.g., M. Castagnino and R. Ferraro, Ann. Phys. (N.Y.) {\bf 154}, 1
(1984); V. V. Dodonov, A. B. Klimov, and D. E. Nikonov,
J. Math. Phys. {\bf 34}, 2742 (1993); C. K. Law, Phys. Rev. Lett. {\bf
73}, 1931 (1994).

\bibitem{Yab}
E. Yablonovitch, Phys. Rev. Lett. {\bf 62}, 1742 (1989).

\bibitem{Dodonov93PRA}
V. V. Dodonov, A. B. Klimov, and D. E. Nikonov, Phys. Rev. A {\bf 47},
4422 (1993).

\bibitem{Johnston}
H. Johnston and S. Sarkar, Phys. Rev. A {\bf 51}, 4109 (1995).

\bibitem{Saito}
H. Saito and H. Hyuga, J. Phys. Soc. Jpn. {\bf 65}, 1139 (1996); {\em
ibid.} {\bf 65} 3513 (1996).

\bibitem{Dodonov90}
V. V. Dodonov, A. B. Klimov, and V. I. Man'ko, Phys. Lett. A {\bf 149},
225 (1990).

\bibitem{Jaekel92}
M. T. Jaekel and S. Reynaud, J. Phys. I France {\bf 2}, 149 (1992).

\bibitem{Plunien}
G. Plunien, R. Sch\"{u}tzhold, and G. Soff, Phys. Rev. Lett. {\bf 84},
1882 (2000).

\bibitem{Dalvit}
D. A. R. Dalvit and P. A. Maia Neto, Phys. Rev. Lett. {\bf 84},
798 (2000).

\bibitem{Razavy83}
M. Razavy, Lett. Nuovo Cimento {\bf 37}, 449 (1983).

\bibitem{Barton}
G. Barton and C. Eberlein, Ann. Phys. (N.Y.) {\bf 227}, 222 (1993).

\bibitem{Law94}
C. K. Law, Phys. Rev. A {\bf 49}, 433 (1994).

\bibitem{Haro}
J. Haro and E. Elizalde, Phys. Rev. Lett. {\bf 97}, 130401 (2006).

\bibitem{Saito02}
H. Saito and H. Hyuga, Phys. Rev. A {\bf 65}, 053804 (2002).

\bibitem{Dodonov96}
V. V. Dodonov and A. B. Klimov, Phys. Rev. A {\bf 53}, 2664 (1996).

\bibitem{Braggio}
C. Braggio, G. Bressi, G. Carugno, C. Del Noce, G. Galeazzi, A. Lombardi,
A. Palmieri, G. Ruoso, and D. Zanello, Europhys. Lett. {\bf 70}, 754
(2005).

\bibitem{Higbie}
J. M. Higbie, L. E. Sadler, S. Inouye, A. P. Chikkatur, S. R. Leslie,
K. L. Moore, V. Savalli, and D. M. Stamper-Kurn, Phys. Rev. Lett. {\bf
95}, 050401 (2005).

\bibitem{Castin}
Y. Castin and R. Dum, Phys. Rev. Lett. {\bf 79}, 3553 (1997).

\bibitem{Law02}
C. K. Law, P. T. Leung, and M. -C. Chu, Phys. Rev. A {\bf 66}, 033605
(2002).

\bibitem{Calzetta}
E. A. Calzetta and B. L. Hu, Phys. Rev. A {\bf 68}, 043625 (2003).

\bibitem{Ho}
T. -L. Ho, Phys. Rev. Lett. {\bf 81}, 742 (1998).

\bibitem{Ohmi}
T. Ohmi and K. Machida, J. Phys. Soc. Jpn. {\bf 67}, 1822 (1998).

\bibitem{Pethick}
See, e.g., C. J. Pethick and H. Smith, {\it Bose-Einstein Condensation in
Dilute Gases} (Cambridge Univ. Press, Cambridge, 2002).

\bibitem{Stenger}
J. Stenger, S. Inouye, D. M. Stamper-Kurn, H. -J. Miesner,
A. P. Chikkatur, and W. Ketterle, Nature (London) {\bf 396}, 345 (1998).

\bibitem{Saito05}
H. Saito and M. Ueda, Phys. Rev. A {\bf 72}, 023610 (2005).

\bibitem{SaitoL}
H. Saito, Y. Kawaguchi, and M. Ueda, Phys. Rev. Lett. {\bf 96}, 065302
(2006); Phys. Rev. A {\bf 75}, 013621 (2007).

\bibitem{SaitoKZ}
H. Saito, Y. Kawaguchi, and M. Ueda, Phys. Rev. A {\bf 76}, 043613
(2007).

\bibitem{Sadler}
L. E. Sadler, J. M. Higbie, S. R. Leslie, M. Vengalattore, and
D. M. Stamper-Kurn, Nature (London) {\bf 443}, 312 (2006).

\bibitem{Black}
A. T. Black, E. Gomez, L. D. Turner, S. Jung, and P. D. Lett,
Phys. Rev. Lett. {\bf 99}, 070403 (2007).

\bibitem{Damski}
B. Damski and W. H. Zurek, Phys. Rev. Lett. {\bf 99}, 130402 (2007).

\end{thebibliography}
\end{document}